\documentclass[twocolumn,english,showpacs,pra]{revtex4}
\usepackage[T1]{fontenc}
\usepackage[latin9]{inputenc}
\usepackage{amsmath}
\usepackage{amssymb}
\usepackage{graphicx}
\usepackage{esint}

\makeatletter

\providecommand{\tabularnewline}{\\}

\@ifundefined{textcolor}{}
{%
 \definecolor{BLACK}{gray}{0}
 \definecolor{WHITE}{gray}{1}
 \definecolor{RED}{rgb}{1,0,0}
 \definecolor{GREEN}{rgb}{0,1,0}
 \definecolor{BLUE}{rgb}{0,0,1}
 \definecolor{CYAN}{cmyk}{1,0,0,0}
 \definecolor{MAGENTA}{cmyk}{0,1,0,0}
 \definecolor{YELLOW}{cmyk}{0,0,1,0}
}

\providecommand{\tabularnewline}{\\}

\@ifundefined{textcolor}{}{%
 \definecolor{BLACK}{gray}{0}
 \definecolor{WHITE}{gray}{1}
 \definecolor{RED}{rgb}{1,0,0}
 \definecolor{GREEN}{rgb}{0,1,0}
 \definecolor{BLUE}{rgb}{0,0,1}
 \definecolor{CYAN}{cmyk}{1,0,0,0}
 \definecolor{MAGENTA}{cmyk}{0,1,0,0}
 \definecolor{YELLOW}{cmyk}{0,0,1,0}
}

\usepackage{amsfonts}

\usepackage{babel}

\usepackage{babel}

\makeatother

\usepackage{babel}
\begin{document}

\title{Polarization properties of solid-state organic lasers}

\author{I. Gozhyk$^{1,*}$, G. Clavier$^{2}$, R. M\'{e}allet-Renault$^{2}$,
M. Dvorko$^{2}$, R. Pansu$^{2}$, J.-F. Audibert$^{2}$, A. Brosseau$^{2}$,
C. Lafargue$^{1}$, V. Tsvirkun$^{1}$, S. Lozenko$^{1}$, S. Forget$^{3}$,
S.Ch\'{e}nais$^{3}$, C. Ulysse$^{4}$, J. Zyss$^{1}$ and M. Lebental$^{1}$}

\affiliation{$^{1}$ Laboratoire de Photonique Quantique et Mol\'{e}culaire, CNRS
UMR 8537, Institut d'Alembert FR 3242, Ecole Normale Sup\'{e}rieure de
Cachan, 61 avenue du pr\'{e}sident Wilson, F-94235 Cachan, France.\\
 $^{2}$ Laboratoire de Photophysique et Photochimie Supramol\'{e}culaires
et Macromol\'{e}culaires, CNRS UMR 8531, Institut d'Alembert FR 3242,
Ecole Normale Sup\'{e}rieure de Cachan, F-94235 Cachan, France.\\
 $^{3}$ Universit\'{e} Paris 13, Sorbonne Paris Cit\'{e}, Laboratoire de
Physique des Lasers, CNRS UMR 7538, F-93430, Villetaneuse, France.\\
 $^{4}$Laboratoire de Photonique et Nanostructures, CNRS UPR20, Route
de Nozay, F-91460 Marcoussis, France. }
\begin{abstract}
The polarization states of lasers are crucial issues both for practical
applications and fundamental research. In general, they depend in
a combined manner on the properties of the gain material and on the
structure of the electromagnetic modes. In this paper, we address
this issue in the case of solid-state organic lasers, a technology
which enables to vary independently gain and mode properties. Different
kinds of resonators are investigated: in-plane micro-resonators with
Fabry-Perot, square, pentagon, stadium, disk, and kite shapes, and
external vertical resonators. The degree of polarization $P$ is measured
in each case. It is shown that although TE modes prevail generally
($P$>0), kite-shaped micro-laser generates negative values for $P$,
i.e. a flip of the dominant polarization which becomes mostly TM polarized.
In general, we demonstrate that both the pump polarization and the
resonator geometry can be used to tailor the polarization of organic
lasers. With this aim in view, we, at last, investigate two other
degrees of freedom, namely upon using resonant energy transfer (RET)
and upon pumping the laser dye to a higher excited state. We then
demonstrate that significantly lower $P$ factors can be obtained. 
\end{abstract}

\pacs{42.55.Sa, 42.55.Mv, 05.45.Mt, 03.65.Yz, 42.60.Da}

\maketitle

\section{Introduction}

Light-matter coupling issues are firmly based on quantum electrodynamics
foundations. However, practical consequences on real systems are often
difficult to derive due to sometimes complicated quantum formulations.
Maxwell-Bloch equations provide a semi-classical expression more appropriate
for lasers \cite{tureci}, which are usually macro- or mesoscopic
systems. The resulting non-linear coupled equations could be handled
by means of numerical simulations, which nevertheless face major problems
for large systems due to huge meshes. This obstacle is even increased
when polarization states of electromagnetic modes are involved, since
they require the treatment of three dimensional and vectorial Maxwell
equations. And yet, polarization remains a key point for many photonics
components.\\

We would like to address this issue by way of micron- and millimeter-sized
lasers of various resonator geometries, which are out of reach of
full electromagnetic simulations due to their large scale, but where
validity of the semi-classical (or geometrical optics) limit is expected
to provide a simplified insight \cite{bogomolny}. In this work, we
propose to evidence and analyze polarization effects in solid-state
organic laser systems, and demonstrate the possibility to modify the
out-put polarization by playing on cavity shape or on material related
features.\\
 Previous works have been devoted to micro-resonators of circular
geometry \cite{frateschi,kim,tsujimoto}, coated fibers \cite{frolov,wang},
and distributed feedback lasers (see for instance \cite{ye}). In
this work, we focus on vertical emitting devices 'VECSOL' (Vertical
External Cavity Surface Emitting Organic Laser \cite{Hadi1}), where
the properties of the gain material can be quite easily decoupled
from the cavity shape, and on thin-film planar micro-lasers of various
contours, such as square, pentagon, kite,... Actually planar micro-resonators
have become widely used in photonics systems, from integrated optics
\cite{masko} to fundamental physics (see for instance \cite{favero}
or \cite{kippenberg}). But in general, their use is limited to Fabry-Perot
(i.e. the resonance occurs between flat parallel edges) or circular
shapes, namely spheres, disks, rings, or tori, while a great variety
of geometries (polygons, stadium, etc...) can be easily fabricated
with nanometric etching quality, providing specific advantages, such
as a higher directivity of emission \cite{djellali,capasso}, a better
coupling to waveguides \cite{poon}, or a high stability of modes
versus perturbations \cite{double-diamant}. The studies of these
geometrical features remain to be improved - in particular when the
polarization of modes is involved - and should lead to optimized devices
for both fundamental and applied photonics. In particular, we will
show hereafter that in-plane polarization is in general favored by
gain and propagation, whereas out-of-plane polarization could be of
crucial importance for applications, like sensing \cite{vollmer}
for instance. We will then propose different ways to monitor the ratio
of polarizations, making use either of the gain or of the resonator
shape.\\

The experiments were carried out with solid-state organic lasers \cite{sebs_polymere}.
Their interest for this study is twofold. Firstly, organic technology
ensures a high etching quality from a relatively fast and easy fabrication,
which enables to investigate a great variety of resonator shapes.
Secondly, the flexibility of organic chemistry allows to explore various
gain media in different pumping schemes in order to monitor the polarization
states, as demonstrated in Sec. \ref{sec:fret} and \ref{sec:Sn}.
The analysis of the experimental results is then performed in the
framework of 'polarization spectroscopy', a domain specific to organic
materials, which matured in the late 80's and has given birth to various
applications in polymer physics or biology (see \cite{lakowicz} and
\cite{valeur} for a review). This domain is based on the fluorescence
anisotropy of dye molecules. A few theoretical articles \cite{casperson1,casperson2,yaroshenko}
extend its range to the non-linear regime of stimulated emission and
lasers, as it was soon evidenced that unlike most other solid-state
laser media, the output polarization of a solid-state dye laser does
not only depend on the anisotropy of the gain medium or on the polarization-dependant
losses due to the cavity, but also depends on the pump beam polarization
\cite{casperson1,casperson2}. The objective of this paper is to better
understand the polarization characteristics of planar organic micro-resonators,
in order to tailor the polarization output of these devices.\\

The paper is organized as follows. The experimental configurations
are described in Sec.\ref{sec:set-up}. Basics on fluorescence anisotropy
are then recalled in Sec.\ref{sec:basics-aniso}, a more detailed
analysis being postponed to the appendixes. The specific case of amplified
spontaneous emission (ASE) is dealt with in Sec.\ref{sec:ASE}. Then
polarizations of laser modes are reported and discussed for various
cavity shapes in Sec.\ref{sec:formes}. These results evidence a strong
influence of the polarization of the pump beam as expected, which
prevents to populate numerous families of modes, that would be otherwise
available. In order to release this constraint and improve the accessibility
of modes, we propose and demonstrate two different methods to tailor
the polarization of the emission: on the one hand, a non-radiative
energy transfer from an excited molecule to the dye laser (Sec.\ref{sec:fret})
and, on the other hand, the direct pumping to a higher excited state
of the dye molecule (Sec.\ref{sec:Sn}).\\

\section{Experimental set-ups}

\label{sec:set-up}

\begin{figure}[htb]
\centerline{\includegraphics[width=8.3cm]{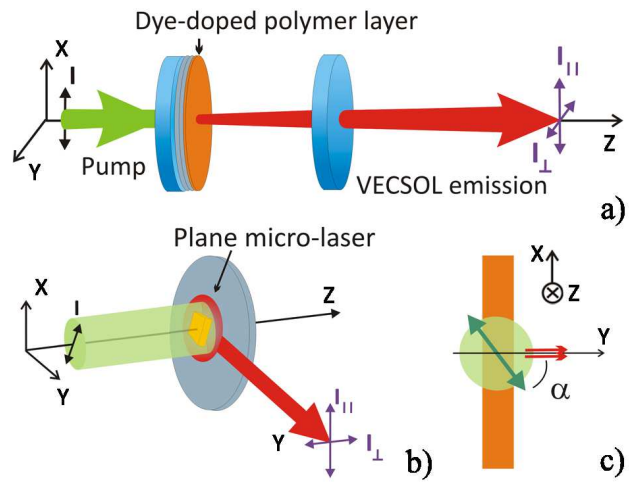}} \caption{(Color
  online) Experimental configurations of the solid-state organic lasers: a)
vertical emission (VECSOL) and b) planar micro-laser (oblique view);
c) Planar micro-laser geometry from a top view ($xy$ plane), indicating
the $\alpha$ angle between the polarization of the pump beam and
the direction of observation ($y$), illustrated with the case of
a Fabry-Perot cavity. For a,b, and c, the colors are green for pumping
and red for emission.}

\label{fig:set-up} 
\end{figure}

\begin{figure}[htb]
\centerline{\includegraphics[width=8.3cm]{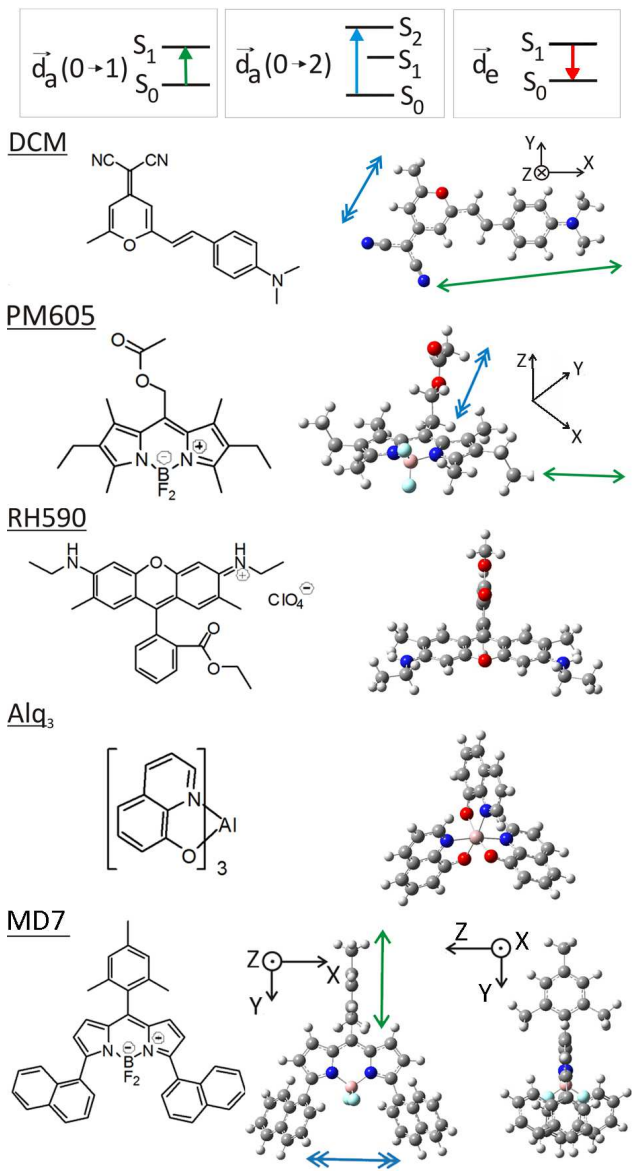}} \caption{(Color
  online) The molecular energy levels involved in the transitions are schematically
shown on the top of the Figure. Dye molecules involved in the study:
DCM, PM605, RH590, Alq$_{3}$, and MD7. MD7 is a home-made modified
pyrromethene. For DCM, PM605, and MD7 the calculated absorption transition
dipoles are indicated with a single arrow for the S$_{0}\rightarrow$S$_{1}$
transition and with a double arrow for the S$_{0}\rightarrow$S$_{2}$
transition. For PM605, two transitions with similar oscillator strengths
are involved when pumped at 355 nm. One of the absorption dipole is
parallel to the absorption dipole of the S$_{0}\rightarrow$S$_{1}$
transition and is thus not represented.}

\label{fig:dipole} 
\end{figure}

\begin{figure}[htb]
\centerline{\includegraphics[width=8.3cm]{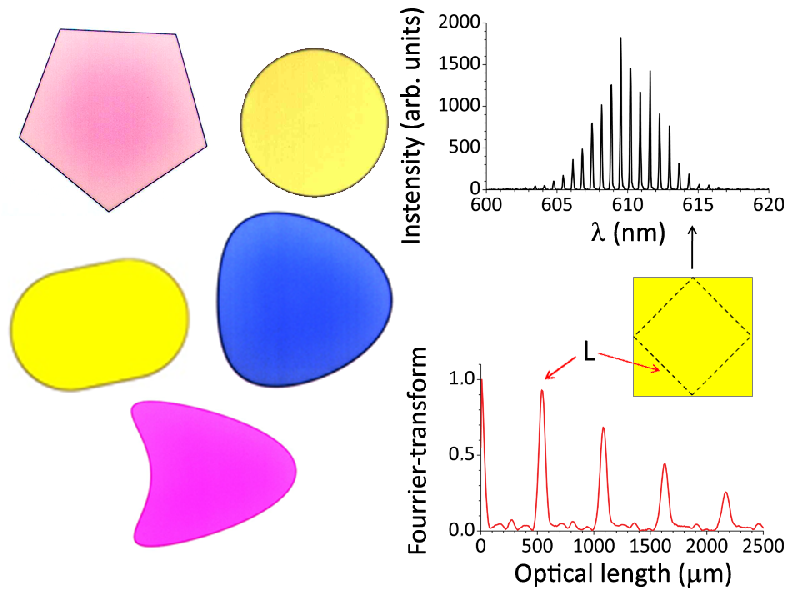}}\caption{(Color
  online) (a) Optical microscope photos of planar microlasers, which are investigated
in this paper. Scales are 0.6 $\mu$m for the thickness and about
100 $\mu$m in-plane. Colors depend on the dye molecules. (b) Typical
experimental spectrum of a square micro-laser. (c) Fourier transform
of the spectrum in (b). The position of the first peak indicates the
optical length $L$ of the periodic orbit (i.e. diamond) drawn in
the inset, according to the data process described in \cite{lebental-spectre}.}

\label{fig:photos} 
\end{figure}

In this paper, we consider thin-film based lasers in two different
configurations represented in Fig.\ref{fig:set-up}. The gain layer
is made of a spin-coated poly(methylmethacrylate) (PMMA) film doped
with laser dyes %
\footnote{In this paper, each commercial dye was bought from Exciton, and PMMA
from MicroChem, 6$\%$ w.t. in anisole 495 000 average chain-length
for in-plane micro-lasers and 15\% w.t. 950 000 for VECSOL.%
}, which are either commercial molecules, namely DCM, pyrromethene
605 (PM605), and rhodamine 590 (RH590), or non-commercial dye like
MD7 \cite{MD7}. Their molecular structures are presented in Fig.\ref{fig:dipole}.
To optimize the lasing efficiency, the concentration is typically
5$\%$ wt for in-plane micro-lasers and 1$\%$ wt for VECSOL, and
the layer thickness is 0.6 $\mu$m and 20 $\mu$m, respectively. In
case of VECSOL, the substrate is directly the back mirror of the cavity,
while for in-plane micro-resonators it is a commercial silicon wafer
with a 2 $\mu$m silica buffer layer.\\
 The specificities of each device are depicted in Fig.\ref{fig:set-up}.
In VECSOL (Fig.\ref{fig:set-up}a), the cavity feedback is ensured
by a curved dielectric mirror \cite{Hadi1}. The pump beam radius
is matched to the fundamental TEM00 cavity mode, which is much smaller
than the diameter of the mirrors. So we assume a rotational symmetry
of the set-up, which is only broken by the polarization of the pump
beam. In this sense, this geometry enables studying the sole influence
of lasing gain medium on polarization, irrespective of any cavity-related
effect. On the contrary, it is the geometries of the in-plane micro-lasers
(Fig.\ref{fig:set-up}b) which determines the types of modes which
are lasing. Such cavities are fabricated from the single gain layer
by electron-beam lithography, which ensures nanometric etching quality
\cite{lebental-matsko}. Arbitrary cavity contours can be designed
(see Fig.\ref{fig:photos}a) to act as resonators. The emission of
a single cavity is then collected in its plane.\\
 Both types of devices are pumped with a pulsed linearly polarized
frequency-doubled Nd:YAG laser (532 nm, 500 ps, 10 Hz). The emission
is injected via a fiber to a spectrometer connected to a cooled CCD
camera, allowing to infer the lasing modes from their spectrum (see
Fig.\ref{fig:photos}b and c) \cite{lebental-spectre}. A polarizer
is set between the device under study and the fiber to project the
electric field of the far-field emitted beam onto two orthogonal directions,
called $I_{||}$ and $I_{\perp}$ (see Fig.\ref{fig:set-up}).\\
 The VECSOL configuration is close to the usual geometry in fluorescence
anisotropy measurements, since $I_{||}$ is registered in the direction
parallel to the pump polarization and $I_{\perp}$ in the orthogonal
direction. On the contrary, the case of in-plane micro-lasers is quite
different. Actually the pump beam propagates perpendicularly to the
cavity plane and its size is much larger than a single cavity, so
that the pump intensity may be considered constant over one resonator.
Thus although the polarization of the pump beam always lies within
the cavity plane, the emission may be collected along any line within
the cavity plane, which means that $I_{||}$ is not always parallel
to the pump polarization. However, we uniformize the terminology for
the two configurations by calling $I_{||}$ and $I_{\perp}$ as well,
the components polarized within ($I_{||}$) and perpendicularly ($I_{\perp}$)
to the film plane (Fig.\ref{fig:set-up}b). In order to describe the
orientation of the pump polarization within the cavity plane, we also
introduce the angle $\alpha$ defined as the angle between the pump
polarization and the direction of observation (see Fig.\ref{fig:set-up}c).\\

\section{Basics in emission anisotropy}

\label{sec:basics-aniso}

Before investigating stimulated emission in the following Sections,
we first review a few basic features in fluorescence anisotropy in
connection with our specific solid-state systems.\\
 An isotropic ensemble of dye molecules is known to emit light with
a non-trivial polarization state. This phenomenon is known as fluorescence
anisotropy and has generated a broad literature (see \cite{lakowicz}
and \cite{valeur} for a review). If the pump laser is linearly polarized,
then the fluorescence emission is not \emph{a priori} equally polarized
along the directions parallel $I_{||}$ and perpendicular $I_{\perp}$
to the polarization of the pump (see Fig.\ref{fig:set-up} for notations).
This anisotropy can be quantified by the degree of polarization %
\footnote{Sometimes, the anisotropy parameter $r=(I_{||}-I_{\perp})/(I_{||}+2I_{\perp})$
is used. But in our experiments, no longitudinal component of the
electric field is expected in the far-field, so the normalized factor
is $I_{||}+I_{\perp}$ and not $I_{||}+2I_{\perp}$.%
}: 
\[
P=\frac{I_{||}-I_{\perp}}{I_{||}+I_{\perp}}
\]
which is zero for equal polarizations, and otherwise remains between
-1 and 1 from a mathematical point of view. However its range is restricted
due to physical limitations as discussed later in this Section.\\
 In experiments, the overall inaccuracy can be estimated to less than
0.05 unit for $P$. The $P$ value can be inferred after integration
over the whole spectrum for each polarization or by considering a
specific mode, both methods leading to almost the same value if a
single mode family is involved. \\

In a liquid solution, the dye molecules are free to rotate. The degree
of polarization $P$ is then zero, except at short delays after the
excitation pulse. Once doped into a rigid polymer matrix, like PMMA,
the fluorophores are not yet able to move, either by thermal or Weigert
effects \cite{dutier1,dutier2}, and $P$ could then be non-zero even
under a stationary pumping. However if the dye concentration is high
enough, F\"{o}rster resonant energy transfer (RET) occurs between nearby
molecules and tends to isotropize the emitted fluorescence under continuous
excitation \cite{lefloch}. However, we noticed that RET is no longer
a limitation to the emission anisotropy in the case of stimulated
emission. This observation could be explained by the difference of
timescales, which is of the order of the fluorescence lifetime (i.e.
ns) for RET \cite{lakowicz,valeur} and could be as short as a few
ps for stimulated emission \cite{bulovic,iryna}. In this paper, we
consider various dye molecules doped in a rigid PMMA matrix in a stimulated
emission regime. We then assume that the fluorophores are not rotating
and moreover that RET does not occur between laser dye molecules under
sub-ns pumping conditions.\\

To deal with lasing, a full non-linear approach should be derived.
Some models were already developed \cite{casperson1,casperson2,yaroshenko},
but as a robust theory is not yet mature, we prefer to resort here
to a phenomenological approach. We assume that the dye molecules can
be modeled as independent emitters, and only consider the influence
of the geometry of the laser cavity. We will show hereafter that in
general this simplified approach is able to capture the main physical
features.\\
 The sample is supposed to be exposed to a linearly polarized light.
The probability that a dye molecule absorbs the pump light is proportional
to $\cos^{2}(\chi)$, where $\chi$ is the angle between the pump
polarization and the absorption transition dipole $\vec{d}_{a}$ of
the molecule. Then the emission of the molecule is assumed to be that
of an emitting transition dipole $\vec{d}_{e}$ in the far-field.
The dipoles $\vec{d}_{a}$ and $\vec{d}_{e}$ depend on the transitions
which are involved in the absorption and subsequent emission processes.
In this paper, we will consider the electronic transitions S$_{0}\rightarrow$S$_{1}$
and S$_{0}\rightarrow$S$_{2}$ for absorption, and S$_{1}\rightarrow$S$_{0}$
for emission (see Fig.\ref{fig:dipole}). Each electronic level S$_{i}$
is broaden by vibrations, which allows to consider dye molecules as
an effective 4-level laser system \cite{svelto}. In general, there
is an angle $\beta$ between $\vec{d}_{a}$ and $\vec{d}_{e}$, which
depends on the molecular structure. A more detailed account is given
in Appendix \ref{sec:dipole} and some $\vec{d}_{a}$ were calculated
with Gaussian$^{\copyright}$ software and reported in Fig.\ref{fig:dipole}.\\

The total electric field is then obtained from the integration over
the orientations of the fluorophores. The distribution of these orientations
is not \emph{a priori} isotropic due to polymer stress \cite{thulstrup}
during spin-coating, as it was reported for $\pi$ conjugated polymers
\cite{tammer,toussaere}. However it was observed that PMMA does not
present an alignment due to spin-coating \cite{agan} and that the
dye molecules embedded into PMMA remain isotropically distributed
\cite{novotny} %
\footnote{The molecular weight of the PMMA used in \cite{agan} is 15 000, while
it is 495 000 in our experiments. It is not specified in \cite{novotny}.
This parameter could be relevant for the organization of the layer
by spin-coating. The case of an anisotropic distribution of fluorophores
was theoretically dealt with in \cite{SPIEnous}.%
}. Hence we only consider an isotropic distribution of dye molecules.
The case of an anisotropic distribution of dyes is theoretically addressed
elsewhere \cite{SPIEnous}, and checked experimentally in \cite{iryna}.

\begin{figure}[htb]
\centerline{\includegraphics[width=8.3cm]{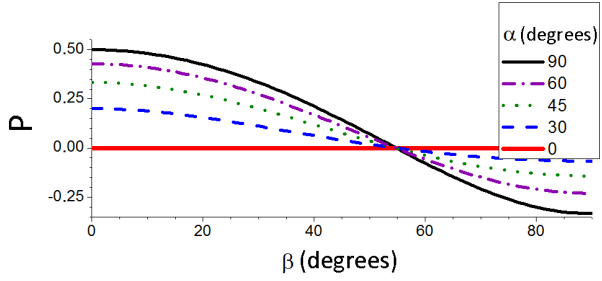}}\caption{(Color
  online) Degree of polarization $P_{in\, plane}$ versus the angle $\beta$
between the absorption and emission dipoles, following expression
(\ref{eq:P-plan}). For $\alpha=\pi/2$, $P_{in\, plane}=P_{vecsol}$.}

\label{fig:P-vs-beta} 
\end{figure}

The analytical calculation of $P$ in both configurations is detailed
in Appendix \ref{sec:calcul-p}. For the VECSOL case, according to
our model, $P$ depends only on the angle $\beta$ between the absorption
and emission transition moments of the dye molecule: 
\begin{equation}
P_{vecsol}=\frac{3\cos^{2}\beta-1}{3+\cos^{2}\beta}\label{eq:P-vecsol}
\end{equation}
For the in-plane geometry, $P$ depends also on the angle $\alpha$
of the pump polarization within the gain layer, and on the specific
resonance which is excited. For the bulk case, we show in Appendix
\ref{sec:calcul-p} that the expression of $P$ is the following:
\begin{equation}
P_{in\, plane}=\frac{\left(3\cos^{2}\beta-1\right)(1-\cos2\alpha)}{7-\cos^{2}\beta-\cos2\alpha\left(3\cos^{2}\beta-1\right)}\label{eq:P-plan}
\end{equation}
It is plotted versus the angle $\beta$ in Fig.\ref{fig:P-vs-beta}
for various angles $\alpha$. The degrees of polarization $P$ in
the in-plane and VECSOL configurations are then not equal, except
for $\alpha=\pi/2$. However the general meaning is similar and the
calculations detailed in Appendix \ref{sec:calcul-p} lead to similar
conclusions for both: $I_{\perp}$ does not depend on the orientation
of the pump polarization $\alpha$, and gets maximum (and $P$ minimum)
when $\beta\rightarrow\pi/2$. The practical case of stimulated emission
in specific resonator geometries is developed in the following Sections.

\section{Amplified spontaneous emission (in planar configuration)}

\label{sec:ASE}

\begin{figure}[htb]
\includegraphics[width=8.3cm]{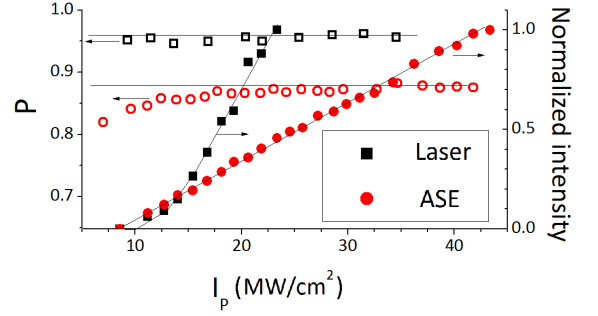}\caption{(Color
  online) Degree of polarization $P$ versus the pump intensity $I_{p}$ (per
pulse) for ASE and Fabry-Perot in planar micro-resonator configuration
for a DCM-PMMA layer. The intensity of emission versus $I_{p}$ is
presented for comparison. Thin lines are drawn for guiding eyes.}

\label{fig:PM605-simple} 
\end{figure}

The emission anisotropy is jointly determined by the molecular properties
and the electromagnetic modes which sustain the generated light. This
Section deals with amplified spontaneous emission (ASE), which involves
the non-linear process of stimulated emission, but does not depend
on the actual resonator shape. In ASE conditions, the emission is
spontaneously generated within the excited gain medium and amplified
by a single path propagation without any feedback.\\
 ASE experiments were carried out in the in-plane configuration, but
without any cavity shaping. An usual DCM-PMMA layer was pumped prior
to any etching and the emission collected in-plane as described in
Sec.\ref{sec:set-up}. Fig.\ref{fig:PM605-simple} (circles) shows
however that $P$ is always higher than the value expected from the
fluorescence model ($P<0.5$ for any angle $\beta$ as seen in Fig.\ref{fig:P-vs-beta}),
and increases with the pump intensity \cite{lam}%
\footnote{At low pump intensities, the ASE data are not shown on Fig.\ref{fig:PM605-simple}
due to low output intensity and thus high experimental uncertainties.%
}. Actually the anisotropy value defined from a fluorescence process
in the previous Section represents the anisotropy well below the ASE
threshold, whenever spontaneous emission is dominant over stimulated
emission. This fluorescence is mainly polarized in plane, since the
pump polarization lies within the plane, and so the dyes oriented
such as $\vec{d}_{a}$ lying in-plane are predominantly excited. And
as $\beta$ is small in general for a $S_{0}$-$S_{1}$ transition,
they predominantly emit in-plane. As the ASE is fed by spontaneous
photons that have a dominant polarization, the excited molecules are
proned to mostly emit photons with this given polarization. Thus avalanche
effect amplifies the difference between the polarized components,
$I_{||}$ is favored \cite{ye}, and then $P$ increases with the
pump intensity.

\begin{figure}[htb]
\includegraphics[width=1\linewidth]{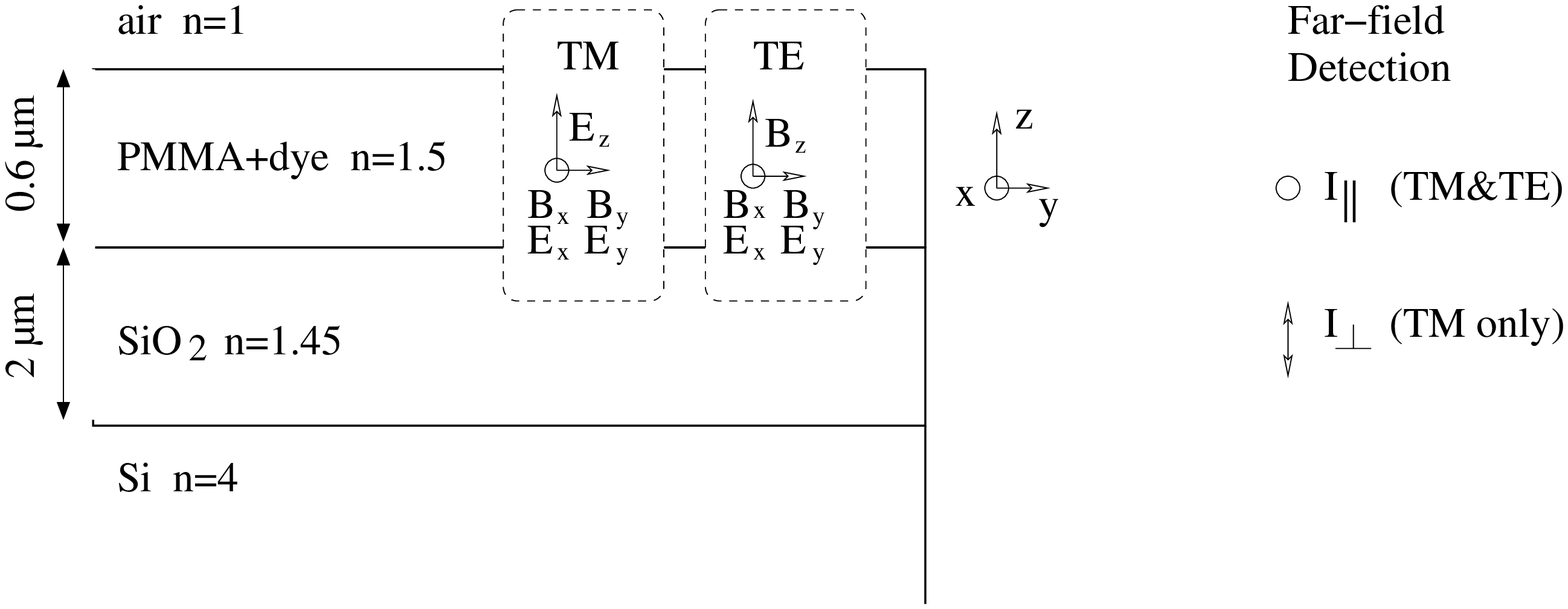}\caption{Scheme of a sample slice in in-plane configuration (not at scale),
with approximative values of bulk refractive indexes at 600 nm.}

\label{fig:ase} 
\end{figure}

The previous discussion summons up the properties of the dye molecules
to explain the prevalence of $I_{||}$ over $I_{\perp}$. Furthermore,
even in the absence of cavity, photons propagate within specific electromagnetic
modes, which also tends to enhance $I_{||}$. Therefore, in order
to get a more comprehensive interpretation, an analysis of the modes
must be performed.\\
 In such in-plane configuration with infinite layers, the approximation
of the effective refractive index applies. The electromagnetic field
can then be split into two independent sets of modes with independent
polarizations, traditionally labeled TE (resp. TM) if there is no
electric (resp. magnetic) component along the $z$ direction (see
Fig.\ref{fig:ase}). It must be pointed out that, as the polarizer
used for analysis is selective on the electric field, a measure of
$I_{\perp}$ is sensitive only to TM modes ($E_{z}$ component), while
$I_{||}$ should \emph{a priori} gather both TE and TM mode contributions
($E_{x}$ component, see Fig.\ref{fig:ase}). In the case of ASE experiments
with infinite layers, the electric field of the TM mode is measured
in the far-field and is thus purely polarized along the $z$ direction
(i.e. no $E_{x}$ component). The strict equivalence TE-$I_{||}$
and TM-$I_{\perp}$ is then valid.\\
 The parameters of our samples are gathered in Fig.\ref{fig:ase}.
The bulk refractive index of PMMA is 1.49 at 600 nm and increased
slightly with the addition of a dye, for instance $n=1.54$ for 5$\%$
wt DCM in PMMA. Assuming an infinite silica layer, then there exists
a single mode for each polarization (one TE and one TM) propagating
inside the doped PMMA layer, both with close effective refractive
index, $n_{eff}\simeq1.50$. However the tiny differences are enhanced
by the non-linearity of stimulated emission. Firstly, the effective
index of TE is slightly higher than that of TM ($\Delta n\simeq5.10^{-3}$),
which means that the TE mode is more localized inside the gain layer
\cite{visser} and thus more amplified. Secondly, the silica layer
is finite. Losses through the silicon layer are then altering mostly
TM mode, which is less confined into the PMMA propagating layer. These
arguments show that mode considerations (without molecular influence)
can explain the discrepancy between both components.\\
 So even for ASE, which is the simplest case involving stimulated
emission, both molecular properties and mode propagation combine to
enhance $I_{||}$, whatever is the dye laser. At low pump intensity,
$P_{DCM}=0.8$ and $P_{PM605}=0.65$. Both $P$ are higher than 0.5,
which is the maximal value expected for an ensemble of isotropically
fluorescent dipoles, for any $\beta$. In the general case of an arbitrary
shape of resonator, the effective index approximation fails at the
boundary (since the layers are not infinite) \cite{bittner}, and
the measure of $P$ may provide an insight into the electromagnetic
modes which co-exist within the cavity as will be considered in the
next Section.\\

\section{Influence of the cavity shape}

\label{sec:formes}

The resonator modifies the degree of polarization in two different
ways. It creates a feedback which enhances further the dominant component
(i.e. $I_{||}$). But at the same time, for in-plane micro-cavities,
reflections at the boundary couple components of the electromagnetic
field and lead to a partial redistribution of the energy.\\
 The simplest case to consider is the Fabry-Perot cavity (i.e. the
classical two-mirrors cavity), in which polarization states have been
extensively studied with liquid dye lasers (see for instance \cite{farland,nagata,yokoyama}),
but more rarely in the solid-state \cite{persano,lam}.\\

\subsubsection*{VECSOLs}

\begin{figure}[htb]
\centerline{\includegraphics[width=8.3cm]{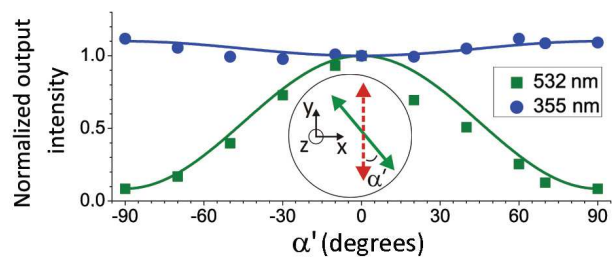}} \caption{(Color
  online) Normalized intensity emitted from a VECSOL with a PMMA-DCM gain material,
versus the angle $\alpha'$ of the pump polarization. Comparison between
a 532 nm pump (green, squares) and a 355 nm pump (blue, circles),
after corrections from bleaching. The curves are fitted according
to Eq. (\ref{eq:courbe-vecsol}). Inset: Scheme of a VECSOL section.
The polarization of emission is fixed and represented as a dotted
red arrow, while the linear pump polarization is shown as a green
continuous arrow.}

\label{fig:VECSOL} 
\end{figure}

In VECSOL configuration with a RH640-PMMA layer, the degree of polarization
$P$ equals to unity for a linearly polarized pump beam, which means
that the lasing emission is totally polarized, like the pump. Moreover
when the pump beam is circularly polarized, the lasing emission is
not polarized, which means that there is no noticeable difference
between the $I_{||}$ and $I_{\perp}$ components, and a quarter waveplate
added on the beam path does not allow to recover a preferential polarization
direction. These results were recorded for a 1 cm-long cavity and
above threshold. According to refs \cite{yokoyama,lam,berg,aiello},
cavity length and pump intensity should be relevant parameters, since
they monitor the build-up time of the modes \cite{hadi2}. Work is
in progress to get a more comprehensive understanding of the polarization
states in this simple geometry which is rotational invariant (axis
$z$, see Fig.\ref{fig:set-up}a).\\
 Another experiment was carried out with a (5\% wt) DCM-PMMA layer,
inserting a glass plate inside the cavity at Brewster angle to force
the emission polarization, and then turning the polarization of the
pump beam by a variable angle $\alpha'$ (see inset of Fig.\ref{fig:VECSOL}
for notation). The laser threshold for $\alpha'=\pi/2$ was found
twice higher than for $\alpha'=0$. Then the pump intensity is fixed
just above the higher threshold and the emitted intensity is recorded
versus $\alpha'$. The results are summarized with squares in Fig.\ref{fig:VECSOL}
and show strong modulations. The curve in Fig.\ref{fig:VECSOL} was
inferred from the calculations presented in Appendix \ref{sec:calcul-p}.
Actually, the geometry of this system is similar to that of the in-plane
configuration with $\alpha=\alpha'$. In the case of fluorescence,
the emitted intensity should then be predicted by Eq. (\ref{eq:I-pour-vecsol}),
which is a linear function of $\cos2\alpha'$. So, 
\begin{equation}
\frac{I(\alpha')}{I(\alpha'=0)}=A+B\cos2\alpha'\label{eq:courbe-vecsol}
\end{equation}
where $\frac{A}{B}=\frac{3+\cos^{2}\beta}{1-3\cos^{2}\beta}$. Eq.(\ref{eq:courbe-vecsol})
is plotted in Fig.\ref{fig:VECSOL} up to a scale parameter and shows
a good agreement with experiments, which indicates that the VECSOL
is working not far from a linear regime, since Eq.(\ref{eq:courbe-vecsol})
is inferred from fluorescence predictions.\\

\subsubsection*{Planar micro-lasers}

In the case of the in-plane configuration, the symmetry is naturally
broken between $I_{||}$ and $I_{\perp}$. Actually TE and TM polarizations
experience slightly different losses during propagation, as mentionned
in the previous Section. Moreover their reflection coefficients at
the micro-resonator boundaries could be different, even at normal
incidence, due to the thinness of the layer \cite{ikegami}. In any
case, $P$ is recorded at a higher level than in ASE experiments,
which indicates that laser feedback further enhances the prevalent
polarization. $P$ depends neither on the specific dye molecule used
($P_{DCM}=P_{PM650}=0.95$), nor on the pump intensity (see Fig.\ref{fig:PM605-simple}),
nor on the cavity length (checked from 100 $\mu$m to 200 $\mu$m).
These observations could be accounted for by a short building time
and short photon lifetime ($\sim$ 1 ps), compared to the fluorescence
lifetime ($\sim$ 1 ns).

\begin{figure}[htb]
\centerline{\includegraphics[width=1\linewidth]{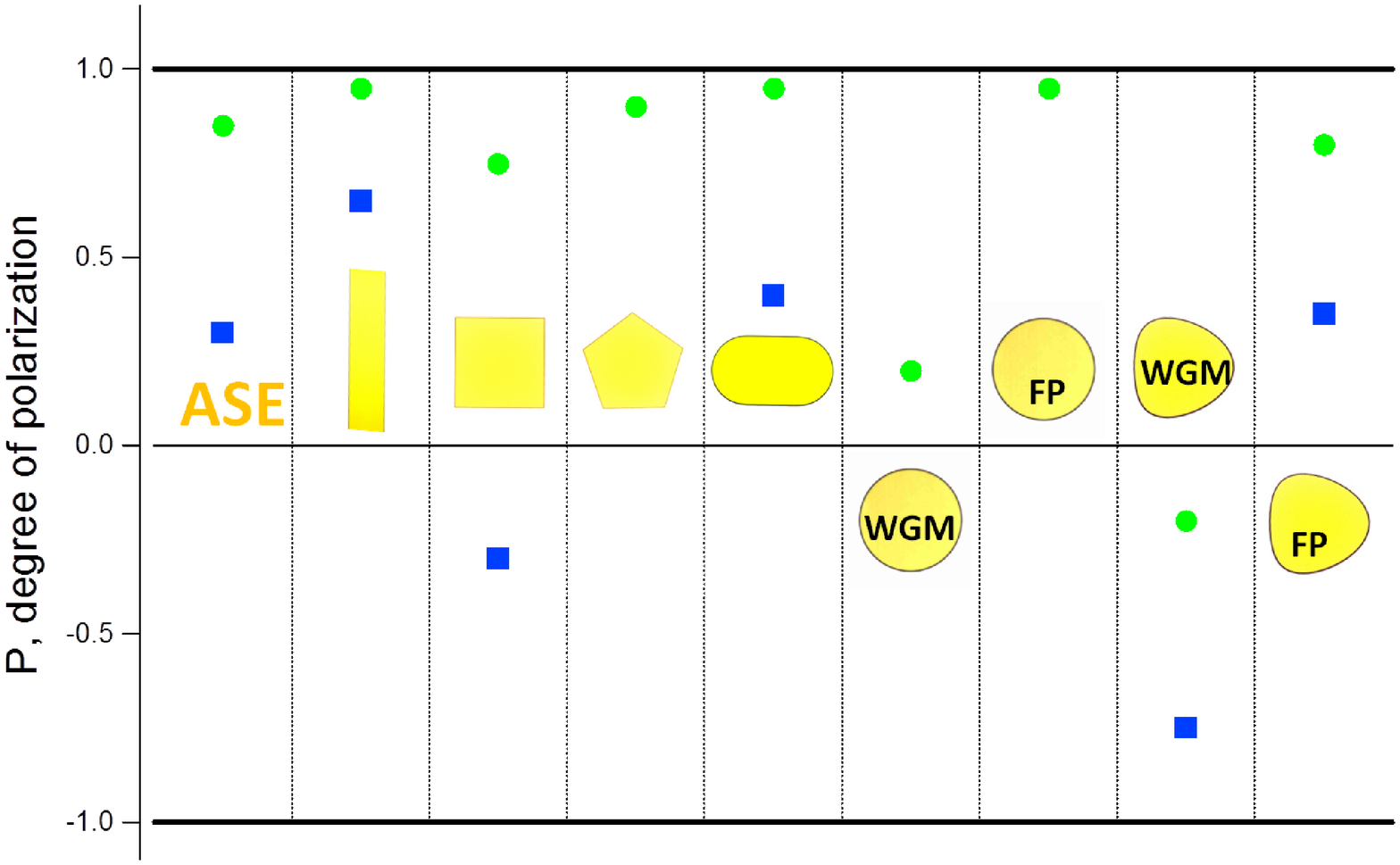}} \caption{(Color
  online) Degree of polarization $P$ in connection with the resonator shape
for a DCM-PMMA layer pumped at 532 nm (green disks) and a MD7-PMMA
layer pumped at 355 nm (blue squares). From left to right: ASE, Fabry-Perot,
square (diamond modes), pentagon (inscribed pentagon), stadium (WGM),
disk (WGM), disk (Fabry-Perot modes), kite (WGM), and kite (Fabry-Perot
modes).}

\label{fig:p-formes} 
\end{figure}

In the in-plane micro-laser configuration, the degree of polarization
$P$ was measured for various cavity shapes and the results gathered
in Fig.\ref{fig:p-formes} and Tab.I. Generally, $P$ depends neither
on $\alpha$, nor on the pump intensity, if recorded high enough above
threshold (typically 20$\%$ higher). The results are quite reproducible,
with error bars about 0.05, which means that the differences between
shapes in Fig.\ref{fig:p-formes} are relevant, and hence due to specific
features of the lasing modes.\\
 In square and pentagon, $P$ is greater than zero, so $I_{||}$ component
dominates. However $P$ is significantly smaller than in a Fabry-Perot
cavity, which evidences a redistribution of the light due to reflections
at the borders. The difference between $P_{square}$ and $P_{pentagon}$
could arise from the periodic orbits sustaining the lasing modes,
namely diamond orbit for square (see Fig.\ref{fig:photos}bc) and
pentagonal orbit for pentagon \cite{lebental-spectre}.\\
 Eventually, the case of whispering gallery modes (WGM) should be
considered. In stadium cavities \cite{lebental-pradirection}, $P$
is close to 1, as in Fabry-Perot lasers, whereas spectral analysis
confirm that the lasing modes are indeed WGM \cite{bogomolny}. Actually
stadium shape leads to chaotic dynamical systems, which could result
in a short photon lifetime ($\sim$ 1 ps)%
\footnote{The photon lifetime of stadiums can be estimated from passive simulations.
In \cite{bogomolny} Fig.23a, the simulation corresponds to the same
shape ratio than the experiments presented here. The imaginary part
of the wavenumber {[}$Im(kR)${]} of the most confined modes seems
to be almost constant versus the real part of the wavenumber. The
photon lifetime can then be estimated from formula $\tau\sim\frac{1}{c\cdot Im(k)}$,
with \textit{c} is a speed of light in vacuum, $Im(kR)\sim0.15$ from
simulations and $R=60$ $\mu m$ in experiments, which leads to $\tau\sim$
1 ps.%
} and then to a lasing behavior close to the Fabry-Perot cavity. With
stadiums, we did not notice any influence on $P$ neither of the pump
polarization $\alpha$, nor of the cavity aspect ratio (ratio between
length and radius, see \cite{bogomolny} for instance).\\
 Disk should be the archetypal shape for WGM. However the presence
of the substrate hinders their observation \cite{lozenko}. The $P$
values reported here were then measured from disks lying on a pedestal.
As reported in \cite{lozenko}, these cavities present two kinds of
lasing mode families: WGM and Fabry-Perot like modes. The later behave
like real Fabry-Perot modes, in particular regarding their $P$ value.
On the contrary, WGM are insensitive to the pump polarization $\alpha$
and their $P$ value is close to zero, probably thanks to a long photon
lifetime%
\footnote{For a perfect 2D disk, the quality factor is huge for the parameter
$kR$, which is used in our experiments ($kR\sim1000$). It is then
difficult to estimate it from numerical calculations. From simulations,
it seems that the quality factor of the best confined modes is growing
logarithmically versus $Re(kR)$: $log_{10}Q\sim0.25Re(kR)$. The extrapolation
to $k=2\pi/0.6$ $\mu m^{-1}$ and $R=100$ $\mu m$ leads then to
$Q\sim10^{250}$, and so to a photon lifetime $\tau=Q/ck\sim10^{235}$
s. Anyway, this photon lifetime is highly shorten by several processes,
such as wall roughness or diffraction at the boundary, and the highest
reported quality factors are about 10\textsuperscript{10} (see a
review in\cite{Vahala}). In our experiments, we expect that the nanometric
quality etching ensures a photon lifetime greater than 10 to 100 ps.%
}, which allows for an efficient mixing of the polarized components
at the boundary \cite{frateschi}. \\
 Finally we considered kite-shaped micro-lasers %
\footnote{The boundary is defined by the polar equation $\rho(t)=\rho_{0}\,(\cos t-2d\, sin^{2}t)$,
with $d=0.165$ in this paper.%
}, which are defined by a slight deformation from a disk \cite{smotrova}
and present the crucial advantage to emit WGM without requiring a
pedestal technology. In that case, $P$ is negative. The structure
of the electromagnetic modes is then allowing by itself to flip the
ratio between the polarized components, which was forced by the gain
properties in the other shapes.\\
 In order to improve the understanding of the mode structure and the
monitoring of the emitted polarization, it would be interesting to
release the prevalence of TE polarization due to the gain material.
As shown in Tab.I, the use of another laser dye does not significantly
alter the TE prevalence. Actually theoretical curves in Fig.\ref{fig:P-vs-beta}
indicate that the angle $\beta$ between the absorption and emission
dipoles should be greater that 55$^{\circ}$ to expect a negative
$P$. As $\beta$ is usually small for S$_{0}$-S$_{1}$ transitions,
a drastic change cannot be expected that way. To significantly improve
$I_{\perp}$ supplies, two different experimental schemes are considered
and implemented in the following Sections.

\begin{table}
\begin{centering}
\begin{tabular}{c|ccc|ccc}
\multicolumn{1}{c}{} & \multicolumn{3}{c}{532 nm} & \multicolumn{3}{c}{355 nm}\tabularnewline
\hline 
{\small 
}\textit{P}  & {\small DCM }  & {\small PM605}  & MD7  & {\small DCM }  & {\small DCM-}Alq$_{3}$  & MD7\tabularnewline
\hline 
ASE  & 0.85  & 0.65  & 0.6  & 0.1  & 0.05  & 0.3\tabularnewline
Fabry-Perot  & 0.95  & 0.95  & 0.9  & 0.9  & 0.9  & 0.65\tabularnewline
Square  & 0.75  & 0.85  & 0.7  & 0.7  & 0.55  & -0.3\tabularnewline
Pentagon  & 0.9  & 0.95  & -  & -  & -  & -\tabularnewline
Stadium  & 0.95  & 0.85  & 0.1  & 0.5  & 0.4  & 0.4\tabularnewline
Disk (FP modes)  & 0.95  & 0.87  & -  & 0.5  & -  & -\tabularnewline
Disk (WGM)  & 0.2  & -  & -  & -  & -  & -\tabularnewline
Kite (FP modes)  & 0.8  & 0.7  & 0.9  & -  & -  & 0.35\tabularnewline
Kite (WGM)  & -0.2  & -0.85  & -0.7  & -0.54  & -0.55  & -0.75\tabularnewline
\hline 
\end{tabular}
\par\end{centering}

{\small \centering \caption{{\small Comparison of }\textit{\small P }{\small obtained from ASE
and various shapes of micro-lasers in the in-plane configuration,
depending on the dye laser and the wavelength of excitation. Some
cavity shapes (e.g. disk and pentagon) were not available for each
laser dye presented in this Table. Moreover lasing under the UV pump
was not achieved systematically for each shape.}}
}{\small \par}

 \label{tab:comparaison} 
\end{table}

\section{Use of resonant energy transfer}

\label{sec:fret}

\begin{figure}[htb]
\centerline{\includegraphics[width=8.3cm]{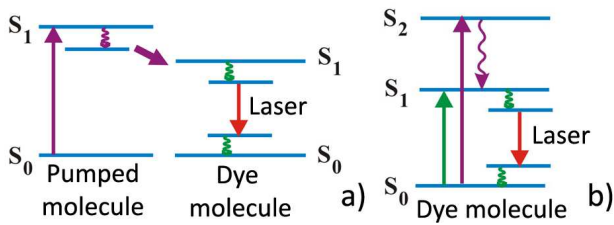}} \caption{(Color
  online) Pumping schemes: (a) via energy transfer from an excited molecule
to the laser dye, (b) direct pumping to the S$_{2}$ state of the
dye laser.}

\label{fig:schema pompe} 
\end{figure}

\begin{figure}[htb]
\centerline{\includegraphics[width=8.3cm]{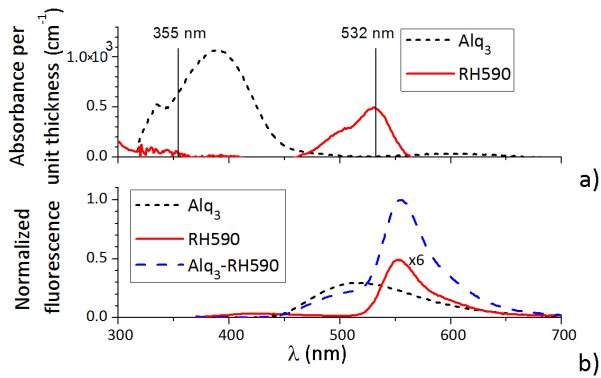}} \caption{(Color
  online) (a) Absorption spectra of Alq$_{3}$ and RH590. (b) Fluorescence spectra
of Alq$_{3}$, RH590 and Alq$_{3}$-RH590 under the same pump intensity
at 355 nm. They are normalized by the maximum of Alq$_{3}$-RH590
fluorescence spectrum and the variation of layer thickness is taken
into account.}

\label{fig:absorption-fluo} 
\end{figure}

An obvious method to promote the $I_{\perp}$ component would be to
use a dye with a molecular structure capable to isotropically redistribute
the pump excitation. For this purpose, the small organo-metallic molecule
Alq$_{3}$ is a good candidate, thanks to its symmetrical propeller
shape. Unfortunately, Alq$_{3}$ cannot be used as a gain material
alone, as in spite of being fluorescent, to our knowledge, no stimulated
emission has been reported to date. But it is possible to add a laser
dye, which will provide stimulated emission after a transfer of excitation
via Alq$_{3}$ \cite{berggren,kozlov}. The scheme of the experiment
is presented in Fig.\ref{fig:schema pompe}a. We expect an emission
in both polarizations due to RET and the specific Alq$_{3}$ geometry.\\
 For this experiment, we used a pulsed frequency-tripled Nd:YAG laser
(355 nm, 300 ps, 10 Hz) to excite the Alq$_{3}$ molecule in its S$_{0}\rightarrow$S$_{1}$
absorption band. For the purity of the demonstration, the required
laser dye should have negligible absorption at the pumping wavelength
(355nm) to ensure that the emission results from energy transfer.
Besides, to provide an efficient energy transfer, the fluorescence
spectrum of Alq$_{3}$ should significantly overlap with the absorption
band of the laser dye. RH590 verifies both criteria as shown in Fig.\ref{fig:absorption-fluo}.\\

\subsubsection*{Planar micro-lasers}

Alq$_{3}$ and RH590 were taken in quantities necessary to satisfy
a 1:1 stoichiometric ratio for 5$\%$ w.t. of RH590 in PMMA. The fluorescence
emission of RH590-Alq$_{3}$ under 355 nm pumping was 10 times more
intense than for RH590 alone (see Fig.\ref{fig:absorption-fluo}b),
which confirms the presence of an efficient energy transfer. In the
case of ASE, the degree of polarization $P$ for RH590 under 532 nm
pumping is 0.5, while introducing Alq$_{3}$ molecules it decreases
to -0.1 under 355 nm pumping. For RH590 alone under 355 nm pumping,
there was even no measurable ASE signal. The transfer of excitation
via Alq$_{3}$ is then an efficient method to increase the participation
of $I_{\perp}$. However the consequences on lasing could not be checked
since no lasing from RH590-Alq$_{3}$ was observed, probably due to
low gain values.\\
 Similar experiments were then performed with DCM, since the transfer
in Alq$_{3}$-DCM is known to be very efficient \cite{kozlov2}. The
perpendicular component of the electric field shows indeed higher
intensities under lasing in various resonator shapes of in-plane micro-lasers.
The $P$ values are summarized in Tab.I. However it is difficult to
say if the decreasing of $P$ can be only assigned to RET via Alq$_{3}$,
since DCM absorbs significantly at 355 nm (see Fig.\ref{fig:fluo-355}a).
Actually we show in the next Section, that absorption to higher excited
states can be used to modify the $P$ values.

\section{Absorption in higher excited states}

\label{sec:Sn}

\begin{figure}[htb]
\centerline{\includegraphics[width=8.3cm]{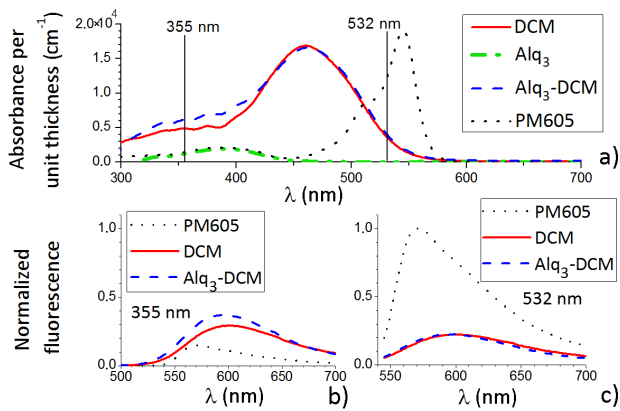}} \caption{(Color
  online) (a) Absorption spectra of PM605, DCM, Alq$_{3}$ and Alq$_{3}$-DCM.
(b-c) Fluorescence spectra of PM605, DCM and Alq$_{3}$-DCM under
pumping at wavelengths of 355 nm (b) and 532 nm (c), normalized by
the maximum of PM605 fluorescence spectrum under 532 nm pumping wavelength.}

\label{fig:fluo-355} 
\end{figure}

To reach small or negative $P$ values - which corresponds to the
involvement of TM polarized modes - Fig.\ref{fig:P-vs-beta} shows
that the angle $\beta$ between the absorption and emission transition
dipoles must be large. However, using the S$_{0}\rightarrow$S$_{1}$
transition only, $\beta$ is usually small. To release this constraint,
a second method is based on pumping the dye laser to a higher excited
states. Actually the absorption dipole of a dye molecule is in general
oriented in a very different manner for S$_{0}\rightarrow$S$_{n}$
(n$>$1). For illustration, the absorption dipoles for S$_{0}\rightarrow$S$_{1}$
and S$_{0}\rightarrow$S$_{2}$ of DCM, PM605, and MD7 were calculated
with Gaussian$^{\copyright}$ software and reported in Fig.\ref{fig:dipole}.
After absorption, it is expected that the molecule relaxes from the
S$_{2}$ state to the S$_{1}$ state, and then emits, as depicted
in Fig.\ref{fig:schema pompe}b. For DCM, PM605, and MD7 some $\beta$
angle between $\vec{d}_{a}(0\rightarrow2)$ and $\vec{d}_{e}(1\rightarrow0)$
are close to $\pi/2$, and negative $P$ values are thus expected,
as predicted in Fig.\ref{fig:P-vs-beta} and reported in \cite{nagata}.
Fig.\ref{fig:fluo-355}b confirms that DCM and PM605 can indeed be
pumped within a higher excited state and then emit from the S$_{1}$
state. It is however difficult to single out which specific state
of the dye lasers is being excited, since the absorption curves of
the transitions partially overlap.\\

Contrary to the case of Sec.\ref{sec:ASE} and in accordance to the
expectations, ASE under 355 nm pumping switches to smaller or negative
$P$ values: $P_{DCM,355}=0.1$ and $P_{PM605,355}=-0.50$ (planar
configuration).\\

\subsubsection*{VECSOLs}

In VECSOL configuration, the experiment of Sec.\ref{sec:formes} was
again carried out, inserting a glass plate inside the cavity at Brewster
angle to force the emission polarization, and then turning the polarization
of the pump beam by an angle $\alpha'$. The results for a DCM-PMMA
layer are plotted with circles in Fig.\ref{fig:VECSOL}. Contrary
to the case of 532 nm pumping, under a high 355 nm pumping, the laser
emission is now almost insensitive to the pump polarization, which
corresponds to $P=0$. Besides, we performed time resolved measurements
of fluorescence anisotropy \cite{pansu} and got $P\sim0$ under a
355 nm pumping, even at very short delay after excitation. This effect
could be assigned to different absorbing transitions, or maybe to
RET occurring during the S$_{n}$ to S$_{1}$ non radiative transitions.
\\

\subsubsection*{Planar micro-lasers}

In micro-laser configuration, pumping into higher excited states is
also an efficient way to modify significantly the ratio between polarized
components. Results for DCM and PM605 are summarized in Tab.I. However
the cavities suffered from a considerable bleaching, whereas it was
not even an issue under the S$_{0}\rightarrow$S$_{1}$ pumping. For
reliable results, we then used a home-made laser dye, called MD7 \cite{MD7},
which molecular structure and absorption dipole moments are presented
in Fig.\ref{fig:dipole}. The degree of polarization $P$ was plotted
in Fig.\ref{fig:p-formes} for various cavity shapes and shows indeed
considerably lower values than under 532 nm pumping. An important
point is that $P$ is still positive for ASE, while it is clearly
negative for diamond modes in square micro-laser, and for WGM in kite
micro-laser. So, although TE polarization is favored due to gain material
and/or propagation, the coupling to TM modes within the resonator
is strong enough to reverse the balance in favor of $I_{\perp}$.
\\
To summarize, with a robust laser dye, pumping into high excited states
is indeed an appropriate way to get a lasing polarization which is
not strictly constrained by the pump polarization and pump geometry.

\section{Conclusion}

In this paper, we investigated the polarization states of organic
solid-state lasers in two different configurations, VECSOL and in-plane
micro-resonators. The framework of fluorescence anisotropy was used
to interpret the data and showed that pump geometry favors a specific
component of the electric field: parallel to the pump polarization
for VECSOL and in-plane for micro-lasers. To release this constraint,
we demonstrated that pumping into higher excited states of the laser
dye can modify significantly the ratio between the polarized components
of the emitted field. These experiments were used to explore the influence
of the resonator shape on the polarization states. For Fabry-Perot
cavities, thanks to the feedback, there is an enhancement of the dominant
polarization compared to amplified spontaneous emission. On the contrary,
for resonances with long photon lifetimes, like WGM, the polarization
states can be strongly modified by coupling of the electromagnetic
components at the cavity boundary. This opens the way to a more systematic
investigation of the relationship between mode structure and resonator
shape, by combining experimental set-up \cite{3D-kim} and numerical
simulations which should be both able to capture the three dimensional
nature of the electromagnetic field.

\section*{Acknowledgments}

The authors acknowledge J. Delaire, S. Brasselet, H. Benisty and S.
Bittner for fruitful discussions, A. Nosich and E. Smotrova for suggesting
kite-shaped cavities, and I. Ledoux-Rak for financial support.

\appendix

\section{Dipolar moments}

\label{sec:dipole}

The absorption transition dipole from the fundamental state $f$ to
an excited state $e$ is defined as the following vector (see p.434
of ref.\cite{braslavski}): 
\begin{equation}
\vec{d}_{a}(f\rightarrow e)=<\psi_{f}\,|\,\hat{\vec{d}}\,|\,\psi_{e}>\label{eq:dipole-abs}
\end{equation}
where $\psi_{f}$ and $\psi_{e}$ are the stationary wavefunctions
of the involved states, and $\hat{\vec{d}}$ is a vector operator
that is the sum of the position vectors of all charged particles weighted
with their charge. Absorption is a fast process ; the positions of
the nucleus are thus assumed to be fixed, and only the electronic
part of the wavefunctions changes between $\psi_{f}$ and $\psi_{e}$.
Laser dyes are often plane aromatic molecules and the S$_{0}\rightarrow$S$_{1}$
transition corresponds to the transfer of a single electron from a
$\pi$ to a $\pi^{\star}$ orbital. These orbitals are both symmetrical
above and below the plane of the molecule, and so the integral (\ref{eq:dipole-abs})
along the direction perpendicular to this plane is zero, since the
global function to integrate is odd. Hence $\vec{d}_{a}(0\rightarrow1)$
lies in the plane of the molecule. If the dye is pumped in its S$_{2}$
state, the excited wavefunction involves in general a $\pi^{\star}$
orbital as well, and $\vec{d}_{a}(0\rightarrow2)$ remains in the
plane of the molecule. However the profile of the $\pi^{\star}$ orbitals
in the plane are different, and thus their corresponding $\vec{d}_{a}$
are not oriented similarly.\\
 The definition of the emission transition dipole is similar to (\ref{eq:dipole-abs}),
except the expression of $\psi_{e}$. Usually the molecule relaxes
before emitting and the wavefunction of the excited state should then
take into account the vibrations of the nucleus. Strictly speaking,
the emission dipole is hence different from the absorption dipole
of the same transition. However the rearrangement in the S$_{1}$
state is in general not very huge and the angle $\beta$ between the
dipoles $\vec{d}_{a}$ and $\vec{d}_{e}$ remains close to zero.\\

Absorption dipoles were calculated with Gaussian$^{\copyright}$ software
and reported in Fig.\ref{fig:dipole}. S$_{0}\rightarrow$S$_{1}$
transition moments are calculated to be (4.04; 0.45; 0) for DCM, (0.3;
-3.1; 0) for MD7 and (2.41; 1.35; 0.17) for PM605, while S$_{0}\rightarrow$S$_{2}$
is (-0.8; -1.27; 0) for DCM, (-1.53; -0.02; -0) for MD7, (-1.06; -0.6;
-0.11) and (0.28; -0.54; 0.2) for PM605. The coordinates correspond
to a frame, which is specific to each molecule and is indicated in
Fig.\ref{fig:dipole}.

\section{Degree of polarization}

\label{sec:calcul-p}

\begin{figure}[htb]
\centerline{\includegraphics[width=8.3cm]{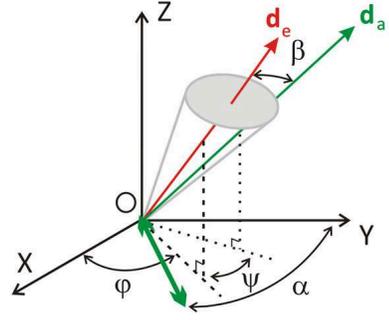}} \caption{(Color
  online) Notations: absorption ($\vec{d}_{a}$) and emission ($\vec{d}_{e}$)
transition dipoles are oriented at the angles $\theta_{a}$ and $\theta_{e}$
with respect to the $z$-axis, $\beta$ - angle between the moments,
$\psi$ - between their projections on the $xy$ plane, $\varphi$
- angle between the $x$-axis and projection of $\vec{d}_{e}$ on
the $xy$-plane, $\alpha$ - angle between the $y$-axis and the pump
beam polarization (thick green arrow).}

\label{fig:angles} 
\end{figure}

In this Appendix, we derive formulas (\ref{eq:P-vecsol}) and (\ref{eq:P-plan})
in a similar way than in \cite{lakowicz} and \cite{valeur}, but
adapted to the specific geometry of the devices in Fig.\ref{fig:set-up}.
Here we assume that the fluorophores are isotropically distributed
in a bulk material and unable to rotate. The case of an anisotropic
distribution is dealt with in \cite{SPIEnous}.\\
 The dye molecules are excited by a linearly polarized electric field
defined by an unit vector $\mathbf{e}=\left\{ \sin\alpha,\,\cos\alpha,\,0\right\} $.
A single molecule from the ensemble is located from its emission transition
dipole $\vec{d}_{e}$ with the usual spherical coordinates $\Omega=(\theta,\varphi)$.
The orientations of its transition dipoles are then characterized
by the following unit vectors 
\begin{eqnarray*}
\mathbf{u_{a}} & = & \frac{\vec{d}_{a}}{||\vec{d}_{a}||}\\
 & = & \left\{ \sin\theta_{a}\cos\left(\varphi+\psi\right),\,\sin\theta_{a}\sin\left(\varphi+\psi\right),\,\cos\theta_{a}\right\} 
\end{eqnarray*}
for absorption and 
\[
\mathbf{u_{e}}=\frac{\vec{d}_{e}}{||\vec{d}_{e}||}=\left\{ \sin\theta\cos\varphi,\,\sin\theta\sin\varphi,\,\cos\theta\right\} 
\]
for emission. The notations are summarized in Fig.\ref{fig:angles}.

We are interested in the emission along axis $z$ for VECSOL configuration
and along axis $y$ for in-plane resonators. In the far-field approximation,
the intensity emitted along a given axis $j$ ($z$ or $y$) can be
presented as a 3D integral of the following product: absorption probability
$P_{a}\propto\left(\mathbf{e}\cdot\mathbf{u_{\mathrm{a}}}\right)^{2}$
and Poynting vector $\Pi_{e\, j}\left(\theta,\varphi\right)\propto\left(\mathbf{j}\times\mathbf{\mathbf{u_{\mathrm{e}}}}\right)^{2}$
integrated over all possible orientations $\Omega$ of the emission
dipoles and $\psi$ of the absorption dipole around the emission dipole:
\begin{equation}
I_{j}=I_{j0}\underset{\Omega}{\int}P_{a}(\Omega)\Pi_{e\, j}(\Omega)d\Omega
\end{equation}

The modulus squared of the Poynting vector can be always presented
as a sum of two orthogonal polarizations: 
\begin{eqnarray*}
\Pi_{e\, y}\left(\Omega\right) & \propto & \underset{I_{||}}{\underbrace{\cos^{2}\theta}}+\underset{I_{\perp}}{\underbrace{\sin^{2}\theta\cos^{2}\varphi}}\\
\Pi_{e\, z}\left(\Omega\right) & \propto & \underset{I_{||}}{\underbrace{\sin^{2}\theta\sin^{2}\left(\varphi+\alpha\right)}}+\underset{I_{\perp}}{\underbrace{\sin^{2}\theta\cos^{2}\left(\varphi+\alpha\right)}}
\end{eqnarray*}
where $I_{||}$ and $I_{\perp}$ are defined in Fig.\ref{fig:set-up}.
After integration over all the possible orientations of the absorption
moment around the emission one, we get the following expression for
the probability of absorption %
\footnote{Error in Eq. (6) of \cite{SPIEnous}. It should be read $\cos2\varphi\cos2\alpha$
instead of $\cos(2\varphi+2\alpha)$%
}: 
\begin{eqnarray}
P_{a}\left(\Omega\right)\propto2-2\cos^{2}\beta & - & (1-3\cos^{2}\beta)\sin^{2}\theta\nonumber \\
 & \times & [1-\cos(2\varphi+2\alpha)]
\end{eqnarray}
Then integration over $\Omega$ must be performed and leads to intensity
components of interest: 
\begin{eqnarray}
I_{||\, in\, plane} & \propto & \frac{1}{30}[3+\cos^{2}\beta+\cos2\alpha(1-3\cos^{2}\beta)]\label{eq:I-pour-vecsol}\\
I_{\perp\, in\, plane} & = & I_{||\, in\, plane}(\alpha=0)\\
I_{||\, vecsol} & = & I_{||\, in\, plane}(\alpha=\pi/2)\nonumber \\
I_{\perp\, vecsol} & = & I_{\perp\, in\, plane}\nonumber 
\end{eqnarray}
Therefore the expressions for the degrees of polarization $P$ are:
\begin{eqnarray}
P_{in\, plane} & = & \frac{\left(3\cos^{2}\beta-1\right)(1-\cos2\alpha)}{7-\cos^{2}\beta-\cos2\alpha\left(3\cos^{2}\beta-1\right)}\\
P_{vecsol} & = & \frac{3\cos^{2}\beta-1}{3+\cos^{2}\beta}
\end{eqnarray}
From these expressions, it follows that for a pump beam polarization
characterized by $\alpha=\pi/2$ (orthogonal to the direction of observation),
then $P_{in\, plane}=P_{vecsol}$. The variation of $P_{in\, plane}$
with $\beta$ angle is depicted on Fig.\ref{fig:P-vs-beta} for several
orientations $\alpha$ of the linear pump polarization.


\begin{thebibliography}{10}
\bibitem{tureci} H. E. Tureci, A. D. Stone, and L. Ge, \pra \textbf{76,}
013813 (2007).

\bibitem{bogomolny} E. Bogomolny, N. Djellali, R. Dubertrand, I.
Gozhyk, M. Lebental, C. Schmit, C. Ulysse, and J. Zyss, \pre \textbf{83,}
036208 (2011).

\bibitem{frateschi} N. Frateschi, A. Kanjamala, A. F. J. Levi, and
T. Tanbun-Ek, \apl \textbf{66,} 1859 (1995).

\bibitem{kim} D. K. Kim, S.-J. An, E. G. Lee, and O'Dae Kwon, J.
Appl. Phys. \textbf{102,} 053104 (2007).

\bibitem{tsujimoto} N. Tsujimoto, T. Takashima, T. Nakao, K. Masuyama,
A. Fujii, and M. Ozaki, J. of Physics D: applied physics, \textbf{40,}
1669 (2007).

\bibitem{frolov} S. V. Frolov, M. Shkunov, Z. V. Vardeny, and K.
Yoshino, \prb \textbf{56,} R4363 (1997).

\bibitem{wang} J. Wang and K. Y. Wong, Appl. Phys. B \textbf{87,}
685 (2007).

\bibitem{ye} Chao Ye, Lei Shi, Jun Wang, Dennis Lo, and Xiao-lei
Zhu, \apl \textbf{83,} 4101 (2003).

\bibitem{Hadi1} H. Rabbani-Haghighi, S. Forget, S. Ch\'{e}nais, and A.
Siove, Opt. Lett. \textbf{35,} 1968 (2010).

\bibitem{masko} \emph{Practical applications of micro-resonators
in optics and photonics}, edited by A. Matsko (CRC, Boca Raton, 2009).

\bibitem{favero} Lu Ding, C. Baker, P. Senellart, A. Lemaitre, S.
Ducci, G. Leo, and I. Favero, \prl \textbf{105,} 263903 (2010).

\bibitem{kippenberg} P. Del'Haye, T. Herr, E. Gavartin, M.L. Gorodetsky,
R. Holzwarth, and T.J. Kippenberg, \prl \textbf{107,} 063901 (2011).

\bibitem{djellali} N. Djellali, I. Gozhyk, D. Owens, S. Lozenko,
M. Lebental, J. Lautru, C. Ulysse, B. Kippelen, and J. Zyss, \apl
\textbf{95,} 101108 (2009).

\bibitem{capasso} Q.J. Wang, C. Yan, N. Yu, J. Unterhinninghofen,
J. Wiersig, C. Pfl\"{u}gl , L. Diehl, T. Edamura, M. Yamanishi, H. Kan,
and F. Capasso, PNAS \textbf{107,} 22407 (2010).

\bibitem{poon} C. Li and A. W. Poon, \ol \textbf{30,} 546 (2005).

\bibitem{double-diamant} See section III.B.2 in \cite{bogomolny}.

\bibitem{vollmer} J. Topolancik and F. Vollmer, Biophysical Journal,
\textbf{92,} 2223 (2007).

\bibitem{sebs_polymere} S. Ch\'{e}nais and S. Forget, Polymer International
\textbf{61,} 390 (2012).

\bibitem{lakowicz} J. R. Lakowicz, \emph{Principles of Fluorescence
Spectroscopy} Springer 2006, 3rd edition

\bibitem{valeur} B. Valeur ``Molecular fluorescence: Principles
and Applications'', 2001, Wiley-VCH

\bibitem{casperson1} K. C. Reyzer and L. W. Casperson, J. Appl. Phys.
\textbf{51,} 6075 (1980).

\bibitem{casperson2} K. C. Reyzer and L. W. Casperson, J. Appl. Phys.
\textbf{51,} 6083 (1980).

\bibitem{yaroshenko} O. I. Yaroshenko, J. Opt. A: Pure Appl. Opt.
\textbf{5,} 328 (2003).

\bibitem{MD7} E. Y. Schmidt, N. V. Zorina, M. Y. Dvorko, N. I. Protsuk,
K. V. Belyaeva, G. Clavier, R. M\'{e}allet-Renault, T. T. Vu, A. B. I.
Mikhaleva, B. A. Trofimov, Chem. Eur. J. \textbf{17,} 3069 (2011).

\bibitem{lebental-matsko} M. Lebental, E. Bogomolny, and J. Zyss,
Organic micro-lasers: a new avenue onto wave chaos physics, in \cite{masko}.

\bibitem{lebental-spectre} M. Lebental, N. Djellali, C. Arnaud, J.-S.
Lauret, J. Zyss, R. Dubertrand, C. Schmit, and E. Bogomolny, \pra
\textbf{76} 023830 (2007).

\bibitem{dutier1} G. Dutier, V. de Beaucoudrey, A. C. Mitus, and
S. Brasselet, Eur. Phys. Lett. \textbf{84,} 67005 (2008).

\bibitem{dutier2} G. Dutier, V. de Beaucoudrey, S. Brasselet, and
J. Zyss, unpublished.

\bibitem{lefloch} V. Le Floc'h, S. Brasselet, J.-F. Roch, and J.
Zyss, J. Phys. Chem. B \textbf{107,} 12403 (2003).

\bibitem{bulovic} V. Bulovi\'{c}, V. G. Kozlov, V. B. Khalfin, and
S. R. Forrest, Science \textbf{279,} 553 (1998).

\bibitem{iryna} I. Gozhyk et al. in preparation.

\bibitem{svelto} O. Svelto, \emph{Principles of Lasers}, Plenum press,
New York (1998).

\bibitem{thulstrup} E. W. Thulstrup and J. Michl, J. Am. Chem. Soc.
\textbf{104,} 5594 (1982).

\bibitem{tammer} M. Tammer and A. P. Monkman, Adv. Mat. \textbf{14,}
210 (2002) ; C. M. Ramsdale and N. C. Greenham, Adv. Mat. \textbf{14,}
212 (2002).

\bibitem{toussaere} J. Sturm, S. Tasch, A. Niko, G. Leising, E. Toussaere,
J. Zyss, T. C. Kowalczyk, K. D. Singer, U. Scherf, and J. Huber, Thin
Solid Films, \textbf{298,} 138 (1997).

\bibitem{agan} S. Agan, F. Ay, A. Kocabas, A. Aydinli, Appl. Phys.
A \textbf{80,} 341 (2005).

\bibitem{novotny} L. Novotny, M. R. Beversluis, K. S. Youngworth,
and T. G. Brown, \prl \textbf{86,} 5251 (2001).

\bibitem{SPIEnous} I. Gozhyk, S. Forget, S. Ch\'{e}nais, C. Ulysse, A.
Brosseau, R. M\'{e}allet-Renault, G. Clavier, R. Pansu, J. Zyss, M. Lebental,
Proceedings of SPIE \textbf{8258,} 82580K (2012).

\bibitem{lam} S. Y. Lam and M. J. Damzen, Appl. Phys. B, \textbf{77,}
577 (2003).

\bibitem{visser} T. D. Visser, B. Demeulenaere, J. Haes, D. Lenstra,
R. Baets, and H. Blok, J. Lightwave Techn. \textbf{14,} 885 (1996).

\bibitem{bittner} S. Bittner, B. Dietz, M. Miski-Oglu, P. O. Iriarte,
A. Richter, and F. Schafer, \pra, \textbf{80,} 023825 (2009).

\bibitem{farland} B. B. McFarland, \apl, \textbf{10,} 208 (1967).

\bibitem{nagata} I. Nagata and T. Nakaya, J. Phys. D: Appl. Phys.
\textbf{6,} 1870 (1973).

\bibitem{yokoyama} X. Wang, R. A. Linke, G. Devlin, and H. Yokoyama,
\pra, \textbf{47,} R2488 (1993).

\bibitem{persano} L. Persano, P. del Carro, E. Mele, R. Cingolani,
D. Pisignano, M. Zavelani-Rossi, S. Longhi, and G. Lanzani, \apl,
\textbf{88,} 121110 (2006).

\bibitem{berg} S. A. van den Berg, V. A. Sautenkov, G. W. 't Hooft,
and E. R. Eliel, \pra, \textbf{65,} 053821 (2002).

\bibitem{aiello} A. Aiello, F. de Martini, and P. Mataloni, Opt.
Lett. \textbf{21,} 149 (1996).

\bibitem{hadi2} H. Rabbani-Haghighi, S. Forget, A. Siove, and S.
Ch\'{e}nais, Eur. Phys. J. Appl. Phys. \textbf{56,} 34108 (2011).

\bibitem{ikegami} T. Ikegami, IEEE J. Quant. Elec. \textbf{8,} 470
(1972).

\bibitem{lebental-pradirection} M. Lebental, J.-S. Lauret, J. Zyss,
C. Schmit, and E. Bogomolny, \emph{Phys. Rev. A} \textbf{75,} 033806
(2007).

\bibitem{lozenko} S. Lozenko, N. Djellali, I. Gozhyk, C. Delezoide,
J. Lautru, C. Ulysse, J. Zyss, and M. Lebental, J. Appl. Phys. \textbf{111,}
103116 (2012).

\bibitem{Vahala} K. J. Vahala, Nature \textbf{424}, 839 (2003).

\bibitem{smotrova} M.V. Balaban, E.I. Smotrova, O.V. Shapoval, V.S.
Bulygin, A.I. Nosich, J. Numerical Modeling: Electronic Networks,
Devices and Fields, vol. 25, 2012, DOI: 10.1002/jnm.1827.

\bibitem{berggren} M. Berggren, A. Dodabalapur, R. E. Slusher, and
Z. Bao, Nature, \textbf{389,} 466 (1997).

\bibitem{kozlov} V. G. Kozlov, V. Bulovi\'{c}, P. E. Burrows, and
S. R. Forrest, Nature, \textbf{389,} 362 (1997).

\bibitem{kozlov2} V. G. Kozlov, V. Bulovi\'{c}, P. E. Burrows, M.
Baldo, V. B. Khalfin, G. Parthasarathy, S. R. Forrest, Y. You, and
M. E. Thompson, J. Appl. Phys. \textbf{84,} 4096 (1998).

\bibitem{pansu} J.-A. Spitz, R. Yasukuni, N. Sandeau, M. Takano,
J.-J. Vachon, R. M\'{e}allet-Renault, and R. B. Pansu, J. of Microscopy-Oxford,
\textbf{229,} 104 (2008).

\bibitem{3D-kim} D. K. Kim, S.-J. An, E. G. Lee, and O'Dae Kwon,
J. Appl. Phys. \textbf{102,} 053104 (2007).

\bibitem{braslavski} E. Braslavsky, \emph{Glossary of terms used
in photochemistry}, Pure Appl. Chem, \textbf{79,} 293 (2007).\end{thebibliography}
\end{document}